%% file: main.tex
\documentclass[runningheads]{llncs}
\usepackage{tipa}   

\usepackage[T1]{fontenc}

\usepackage{tipa}   

\usepackage[T1]{fontenc}
\usepackage{graphicx}
\usepackage{microtype} 
\usepackage{algorithm,algpseudocode}  
\usepackage{xcolor}   

\usepackage{cite}

\usepackage{hyperref}
\usepackage{graphicx}
\usepackage{subcaption}   

\usepackage{wrapfig}
\usepackage{float}
\usepackage{algpseudocode}
\usepackage{amsmath}
\usepackage{amsfonts}  
\usepackage{graphicx}
\usepackage{booktabs}

\usepackage{wrapfig}

\usepackage{booktabs}   
\usepackage{multirow}   
\usepackage{siunitx}    

\usepackage{tikz}                        
\usetikzlibrary{positioning,arrows.meta} 

\title{Pronunciation Deviation Analysis Through Voice Cloning and Acoustic Comparison}

\author{
Andrew Valdivia\inst{1},
Yueming Zhang\inst{1},
Hailu Xu\inst{1},
Amir Ghasemkhani\inst{1},
Xin Qin\inst{1}
}
\institute{
California State University Long Beach, Long Beach, CA, USA
\email{
\{andrew.valdivia01,simon.zhang01\}@student.csulb.edu}
\email{
\{hailu.xu,amir.ghasemkhani,xin.qin\}@csulb.edu
}}

\makeatletter
\newenvironment{breakablealgorithm}{
  \begin{center}
    \refstepcounter{algorithm}
    \hrule height .8pt\kern2pt
    \renewcommand{\caption}[2][\relax]{%
      {\raggedright\textbf{\fname@algorithm~\thealgorithm}\ ##2\par}%
      \addcontentsline{loa}{algorithm}{\protect\numberline{\thealgorithm}##2}%
      \kern2pt\hrule\kern2pt
    }%
}{%
    \kern2pt\hrule
  \end{center}
}
\makeatother

\begin{document}

\maketitle

\begin{abstract}
This paper presents a novel approach for detecting mispronunciations by analyzing deviations between a user's original speech and their voice-cloned counterpart with corrected pronunciation. We hypothesize that regions with maximal acoustic deviation between the original and cloned utterances indicate potential mispronunciations. Our method leverages recent advances in voice cloning to generate a synthetic version of the user's voice with proper pronunciation, then performs frame-by-frame comparisons to identify problematic segments. Experimental results demonstrate the effectiveness of this approach in pinpointing specific pronunciation errors without requiring pre-defined phonetic rules or extensive training data for each target language.
\end{abstract}


\section{Introduction}
Computer-assisted language learning programs have been extensively researched and widely adopted to overcome deficiencies inherent in traditional English language learning resources. These tools enable ESL (English as a Second Language) learners and tutors to systematically address nuanced linguistic challenges, particularly those related to phonological differences between English phonemes and those present in a learner's primary language. In this context, computer-assisted pronunciation training (CAPT) has emerged as a crucial resource, offering learners and instructors accessible methods to practice and refine English pronunciation \cite{amrate2024computer}.

For such systems to be effective, they must robustly detect subtle mispronunciations and provide immediate, actionable feedback to learners. However, existing CAPT systems often underperform due to their reliance on generalized pronunciation models, which fail to account for individual variability in a learner's idiolect and accent characteristics. Standard reference models typically lack personalization, limiting their sensitivity to nuanced pronunciation errors\cite{korzekwa2022computer,richter2023orthography}. Moreover, CAPT systems tend to overlook phonological transfer effects from learners' first languages (L1), which vary widely and require targeted modeling approaches \cite{Korzekwa_2022} .


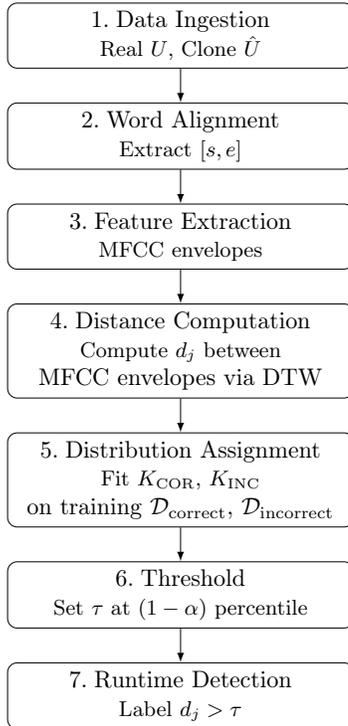
\begin{wrapfigure}{r}{0.42\textwidth}
  \centering
    \resizebox{0.9\linewidth}{!}{%
  \begin{tikzpicture}[
      >=latex,
      node distance=5mm,
      box/.style = {
        draw,
        rounded corners,
        minimum width=5.1cm,
        align=center
      }
    ]

    \node[box] (s1) {1.\;Data Ingestion\\\footnotesize Real $U$, Clone $\hat{U}$};
    \node[box, below=of s1] (s2) {2.\;Word Alignment\\\footnotesize Extract $[s,e]$};
    \node[box, below=of s2] (s3) {3.\;Feature Extraction\\\footnotesize MFCC envelopes};
    \node[box, below=of s3] (s4) {4.\;Distance Computation\\\footnotesize Compute $d_j$ between\\ MFCC envelopes via DTW};
    \node[box, below=of s4] (s5) {5.\;Distribution Assignment \\\footnotesize Fit $K_{\mathrm{COR}}$, $K_{\mathrm{INC}}$\\on training $\mathcal{D}_{\mathrm{correct}}$, $\mathcal{D}_{\mathrm{incorrect}}$};
    \node[box, below=of s5] (s6) {6.\;Threshold\\\footnotesize Set $\tau$ at $(1-\alpha)$ percentile};
    \node[box, below=of s6] (s7) {7.\;Runtime Detection\\\footnotesize Label $d_j > \tau$};

    \foreach \a/\b in {s1/s2, s2/s3, s3/s4, s4/s5, s5/s6, s6/s7}
      \draw[->] (\a.south) -- (\b.north);
  \end{tikzpicture}}
  \caption{Data-processing pipeline.}
  \label{fig:DataPipelineTikz}
  \vspace{-0.8cm}
\end{wrapfigure}

To address these limitations, our approach leverages personalized, synthetically generated voices, finely tuned to replicate each learner's unique vocal traits under ideal pronunciation conditions. We employ advanced voice cloning technologies, explicitly utilizing the ElevenLabs platform \cite{elevenlabs2025}, renowned for its sophisticated deep-learning algorithms and realistic synthetic speech generation. ElevenLabs' state-of-the-art neural models and extensive training datasets enable precise replication of individual vocal nuances, including subtle variations in intonation, rhythm, and articulation. This personalized synthetic reference serves as a tailored benchmark, substantially enhancing mispronunciation detection and feedback mechanisms' sensitivity, accuracy, and real-time responsiveness.

Recent research has shown that integrating speech synthesis into CAPT workflows enables large-scale generation of native-like references for training error detection models, overcoming the scarcity of annotated mispronounced data \cite{Korzekwa_2022} . Furthermore, work by Das and Gutierrez-Osuna \cite{10613461} demonstrated that combining text-to-speech (TTS) with speech reconstruction in a multi-task learning framework improves mispronunciation detection accuracy by modeling the mismatch between learner speech and native-like reconstruction. Nguyen et al. \cite{10890229} extended this idea by synthesizing native-accented versions of non-native speech using knowledge distillation and TTS-based ground truth, allowing more effective accent and pronunciation correction.

Our experiments leverage the L2-ARCTIC corpus \cite{zhao2018l2arctic} a comprehensive dataset designed explicitly for voice conversion, accent modification, and mispronunciation detection research. The corpus includes 26,867 utterances from 24 non-native English speakers representing diverse linguistic backgrounds, such as Arabic, Chinese, Hindi, Korean, Spanish, and Vietnamese. Each speaker provided approximately one hour of phonetically balanced read speech, amounting to over 27 hours of recorded data. Critically, the dataset offers detailed annotations, comprising over 238,000 word segments and approximately 852,000 phone segments, with explicit identification of more than 14,000 phone substitutions, 3,400 deletions, and 1,000 additions. These meticulous annotations facilitate the robust development and evaluation of innovative pronunciation assessment tools.

Integrating personalized voice cloning with a comprehensive, rigorously annotated corpus like L2-ARCTIC enables our method to pinpoint problematic phonemes and subtle pronunciation errors reliably. Consequently, our approach offers targeted, precise feedback and provides more effective pronunciation training interventions. This personalized feedback strategy also aligns with emerging trends in CAPT personalization \cite{khaustova2023capturing},\cite{zhou2022accent}, where learner-specific acoustic profiles, L1 influences, and synthetic benchmarking are used to enhance training efficacy.

\section{Related Work}
Recent advances in voice cloning and neural TTS have reshaped pronunciation training. Modern systems leverage synthetic speech for high-variability phonetic training \cite{khaustova2023capturing}, accent conversion \cite{zhou2022accent}, and personalized feedback generation. Notably, \cite{khaustova2023capturing} demonstrated that synthetic mispronunciations created through phoneme manipulation and voice cloning significantly improve error detection accuracy in CAPT systems, addressing data scarcity in non-native speech corpora. Their work aligns with the broader trend of using generative models to bypass dependency on annotated L2 data – a key enabler for our approach.

Traditional phoneme-level feedback methods rely on acoustic models like GOP scoring or end-to-end neural classifiers \cite{richter2023orthography}. While effective, these techniques require extensive training data and struggle with interpretability, often necessitating post-hoc transformations to map phonetic errors to orthographic representations. Recent innovations in voice identity preservation through neural voice conversion \cite{khaustova2023capturing} suggest new possibilities for creating personalized pronunciation references, though prior work has not exploited acoustic deviation analysis between native and learner-specific voice clones.

Standard CAPT systems face three key limitations our method circumvents: (1) reliance on a large number of samples for model training, (2) rule-based error detection requiring phonetic expertise, and (3) generic feedback lacking voice personalization. By combining voice cloning with frame-level acoustic analysis, our approach eliminates the need for predefined phonetic rules while maintaining native pronunciation benchmarks in the learner’s own vocal characteristics.
\section{Proposed Method}
Our mispronunciation detector runs in two phases, calibration and runtime, over six steps:

\subsection{Data Preparation}
Load real utterances \(U=\{u_i\}\) with TextGrids format and their TTS clones \(\hat U=\{\hat u_i\}\).

    \subsection{Word Alignment}
    For each pair $ (u_i,\hat u_i) $, extract the set of word alignments:
    \begin{equation}
      \bigl\{\, \bigl( w_j, [s_j^{\mathrm{real}}, e_j^{\mathrm{real}}], [s_j^{\mathrm{clone}}, e_j^{\mathrm{clone}}] \bigr) \mid j = 1, \ldots, n_i \,\bigr\},
    \end{equation}
    where $ n_i $ is the number of words in the utterance, $ w_j $ is the $ j $-th word, and $ [s_j^{\mathrm{real}}, e_j^{\mathrm{real}}] $ and $ [s_j^{\mathrm{clone}}, e_j^{\mathrm{clone}}] $ are the start and end times of $ w_j $ in the real and clone utterance, respectively.


\subsection{Feature Extraction}
For each aligned word instance \(j\), compute 13-dimensional MFCC envelopes \cite{itsp2022}:
\begin{align}
E_j^{\mathrm{real}}  &= \left[ \mathbf{e}_{j,1}^{\mathrm{real}},  \mathbf{e}_{j,2}^{\mathrm{real}},  \dots,  \mathbf{e}_{j,T_j^{\mathrm{real}}}^{\mathrm{real}} \right] \in \mathbb{R}^{13 \times T_j^{\mathrm{real}}} \\
E_j^{\mathrm{clone}} &= \left[ \mathbf{e}_{j,1}^{\mathrm{clone}}, \mathbf{e}_{j,2}^{\mathrm{clone}}, \dots, \mathbf{e}_{j,T_j^{\mathrm{clone}}}^{\mathrm{clone}} \right] \in \mathbb{R}^{13 \times T_j^{\mathrm{clone}}}
\end{align}
where \(T_j^{\mathrm{real}}, T_j^{\mathrm{clone}}\) is the number of time frames in real/clone segments respectively, \(\mathbf{e}_{j,t}^{\mathrm{real}}, \mathbf{e}_{j,t}^{\mathrm{clone}} \in \mathbb{R}^{13}\) denotes MFCC feature vectors at frame \(t\).


\subsection{Multi-Feature Dynamic Time Warping}

Given a pair of aligned word instances represented by their 13-dimensional MFCC envelopes, let $A$ denote $E_j^{\mathrm{real}}$ and let $B$ denote $E_j^{\mathrm{clone}}$, where $T_A$ and $T_B$ denote the respective frame counts. We first resample both sequences to a common length $\hat{T} = \max(T_A, T_B)$ along the time axis, yielding $\tilde{A} \in \mathbb{R}^{\hat{T} \times 13}$ and $\tilde{B} \in \mathbb{R}^{\hat{T} \times 13}$. For each MFCC coefficient $d \in \{1,\dots,13\}$, we compute the standard DTW distance $d_d$ between the aligned 1D sequences $\tilde{A}_{:,d}$ and $\tilde{B}_{:,d}$. The final distance $\bar{d} = (13\hat{T})^{-1}\sum_{d=1}^{13} d_d$ averages these per-coefficient distances while normalizing for sequence duration and feature dimensionality. This approach handles temporal misalignments through DTW, accommodates variable-length inputs via resampling, and provides coefficient-level comparisons through the decomposed distances $d_d$, with the normalization ensuring scale invariance across different utterance lengths.

\begin{figure}
  \centering
  \includegraphics[
    width=0.7\textwidth,
    trim=0cm 0cm 0cm 1.5cm,  
    clip                     
  ]{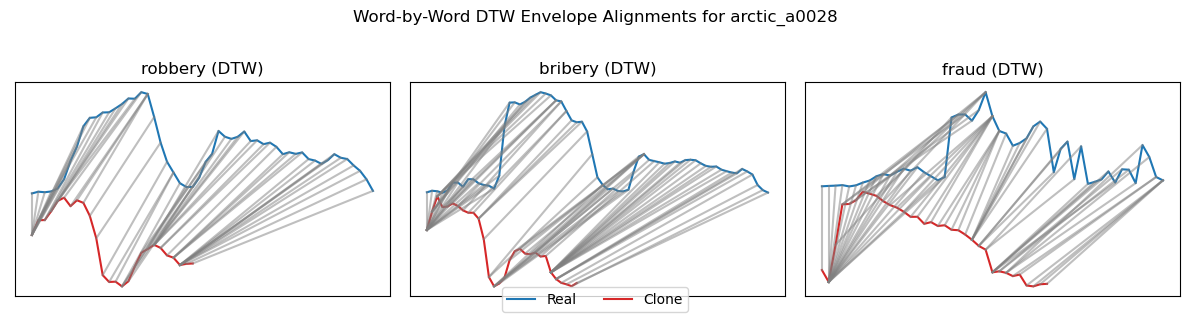}
  \caption{An example of DTW alignment between original and synthesized samples.}
  \label{fig:dtw}
  \vspace{-1cm}
\end{figure}



\subsection{Threshold Calculation}
\label{subsec:threshold_calculation}

Given the sets of training distances partitioned by prediction correctness:
\[
\mathcal{D}_{\mathrm{correct}} = \{\bar{d}_j \mid \text{correct predictions}\}, \quad 
\mathcal{D}_{\mathrm{incorrect}} = \{\bar{d}_j \mid \text{incorrect predictions}\},
\]
the class-specific thresholds $\tau_C$ and $\tau_I$ are computed as the 90th percentiles of their respective distance distributions:

\begin{equation}
\tau_C = Q_{0.9}\left(\mathcal{D}_{\mathrm{correct}}\right), \quad
\tau_I = Q_{0.9}\left(\mathcal{D}_{\mathrm{incorrect}}\right),
\end{equation}
where $Q_{p}(\cdot)$ denotes the $p$-th percentile of a dataset.
\noindent
These thresholds satisfy:
\[
\mathbb{P}(\bar{d}_j \leq \tau_C \mid \text{correct}) = 0.9, \quad
\mathbb{P}(\bar{d}_j \leq \tau_I \mid \text{incorrect}) = 0.9,
\]
where $\mathbb{P}$ denotes the empirical probability.



\subsection{Runtime Detection}
Given the sets of training distances partitioned by prediction correctness $\mathcal{D}_{\mathrm{correct}}$ and $\mathcal{D}_{\mathrm{incorrect}}$, we their probability density functions using kernel density estimation (KDE) \cite{silverman2018density}, and denote with $K_{\mathrm{COR}}(x)$ and $K_{\mathrm{INC}}(x)$, where $K_{\mathrm{COR}}(x) = \text{kde}(\mathcal{D}_{\mathrm{correct}})
$ and $
K_{\mathrm{INC}}(x) = \text{kde}(\mathcal{D}_{\mathrm{incorrect}})$.
We define the decision rule for classification as:
\begin{equation}
    \text{c}(x) = \begin{cases}
0 & \text{if } K_{\mathrm{COR}}(x) \geq K_{\mathrm{INC}}(x) \\
1 & \text{otherwise}
\end{cases}
\end{equation}

For a new utterance \(u\) with synthesized version \(\hat u = V(u)\), we first calculate the distance \(\bar d_j\), then determine the prediction as follows:

The threshold selection is given by:
\begin{equation}
\tau^* = 
\begin{cases}
\tau_C, & \text{if } \mathrm{c}(\bar d_j) = 0,\\
\tau_I, & \text{if } \mathrm{c}(\bar d_j) = 1,
\end{cases}
\end{equation}
The final prediction is assigned according to:
\begin{equation}
\hat y_j =
\begin{cases}
\text{CORRECT},   & \mathrm{c}(\bar d_j) = 0 \land \bar d_j \le \tau_C,\\
\text{INCORRECT}, & \mathrm{c}(\bar d_j) = 1 \land \bar d_j \ge \tau_I,\\
\text{AMBIGUOUS}, & \text{otherwise}.
\end{cases}
\end{equation}

\input{algo_2}

\section{Experiments}
We tested our model on four L2-Arctic speakers (EBVS, ERMS, MBMPS, NJS), measuring classification performance via precision, recall, F1-score, and accuracy.


\begin{table}[htbp]
  \centering
  \begin{tabular*}{\textwidth}{@{\extracolsep{\fill}} l
                               S[table-format=1.0]  
                               S S S S              
                               c                    
                               }
    \toprule
    Dataset & {Class} & {Precision} & {Recall} & {F1-score} & {Support} & {Accuracy} \\
    \midrule
    \multirow{2}{*}{EBVS}   & 0 & 0.615 & 0.582 & 0.598 & 110 & \multirow{2}{*}{0.646} \\
                            & 1 & 0.669 & 0.699 & 0.684 & 133 & \\
    \addlinespace[0.5ex]
    \multirow{2}{*}{ERMS}   & 0 & 0.581 & 0.541 & 0.560 & 133 & \multirow{2}{*}{0.537} \\
                            & 1 & 0.492 & 0.532 & 0.511 & 111 & \\
    \addlinespace[0.5ex]
    \multirow{2}{*}{MBMPS}  & 0 & 0.604 & 0.416 & 0.492 & 154 & \multirow{2}{*}{0.461} \\
                            & 1 & 0.353 & 0.538 & 0.426 &  91 & \\
    \addlinespace[0.5ex]
    \multirow{2}{*}{NJS}    & 0 & 0.634 & 0.479 & 0.545 & 163 & \multirow{2}{*}{0.486} \\
                            & 1 & 0.346 & 0.500 & 0.409 &  90 & \\
    \bottomrule
  \end{tabular*}
    \caption{Classification performance of the proposed model on each dataset.}
  \label{tab:classification}
\end{table}




Table \ref{tab:summary} presents the per-individual distance distribution (mean, median, standard deviation, and 25th/75th percentiles) for correctly and incorrectly classified examples. The results show that mispronounced words exhibit a greater average distance between cloned and original voices compared to correctly pronounced words. Table \ref{tab:classification} shows that across users, our model performs best on EBVS, suggesting the potential for individualized models for each user, which could extend our current methods, which are based on data from all users.

To investigate explainability in our model, we employ the NJS speech sample \texttt{arctic\_a0028.mp2} to illustrate the system’s capacity for resolving an issue that is similar to the caught–cot vowel merger. This phenomenon, stemming from first‐language phonological interference, results in non‐native speakers conflating two distinct English low‐back vowels. English, by contrast, maintains approximately twenty contrastive vowel phonemes, including the low back unrounded (/A/) and low back rounded (/O/) categories.

\begin{wrapfigure}[16]{L}{0.6\linewidth}
  \centering
  \includegraphics[width=0.59\textwidth,trim=2cm 0cm 2cm 0cm,  
    clip]{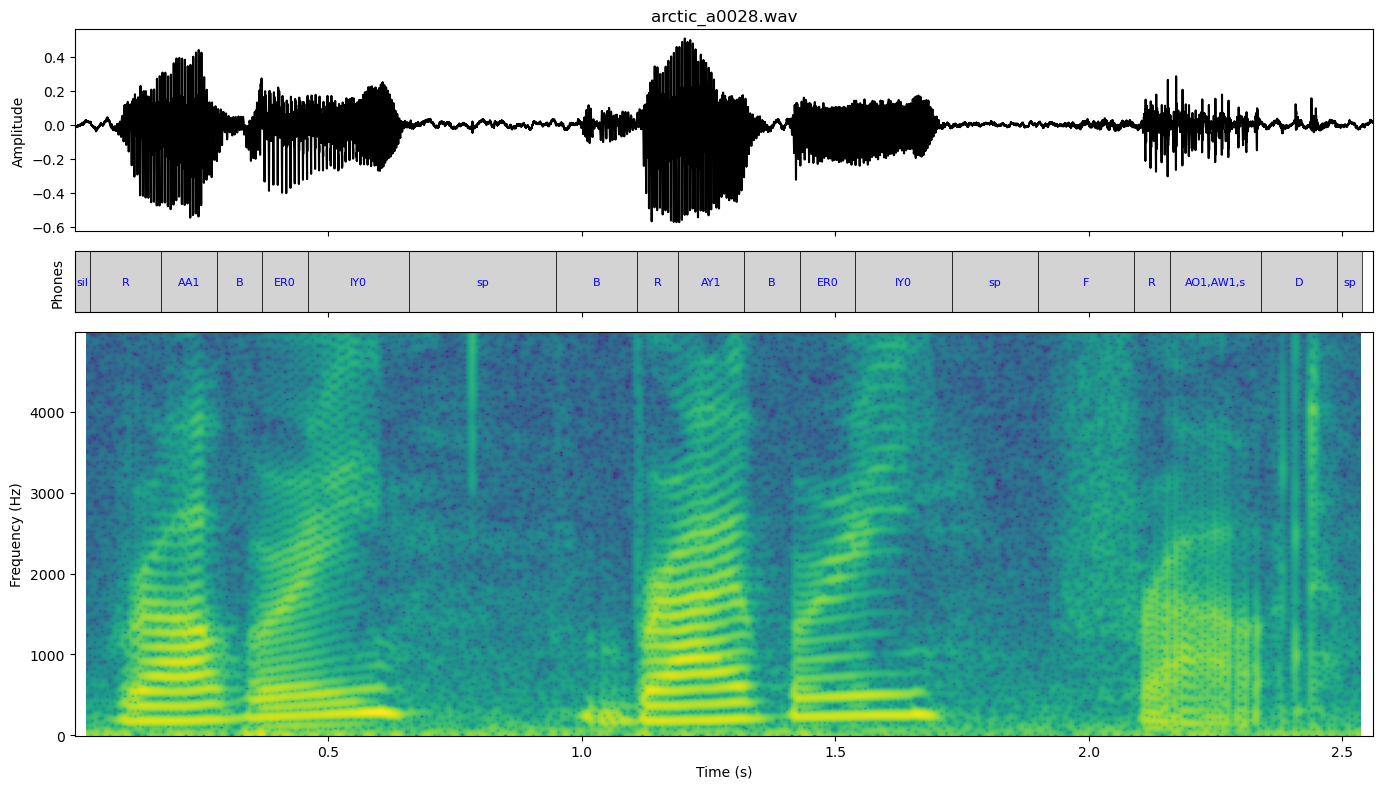}
  \caption{Phonetic transcription of “Robbery, Bribery, Fraud.”}
  \label{fig:phones}
\end{wrapfigure}

\begin{figure}[!t]
  \centering

  \begin{subfigure}[t]{0.32\linewidth}
    \centering
    \includegraphics[width=\linewidth]{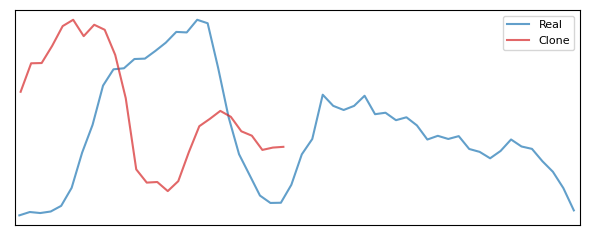}
    \caption{Robbery}
  \end{subfigure}
  \begin{subfigure}[t]{0.32\linewidth}
    \centering
    \includegraphics[width=\linewidth]{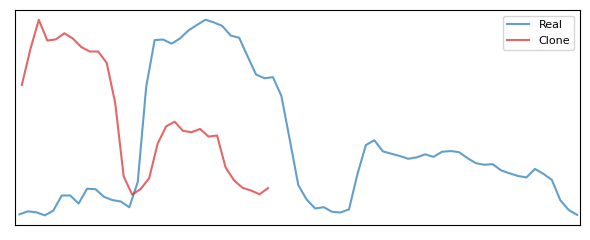}
    \caption{Bribary}
  \end{subfigure}
  \begin{subfigure}[t]{0.32\linewidth}
    \centering
    \includegraphics[width=\linewidth]{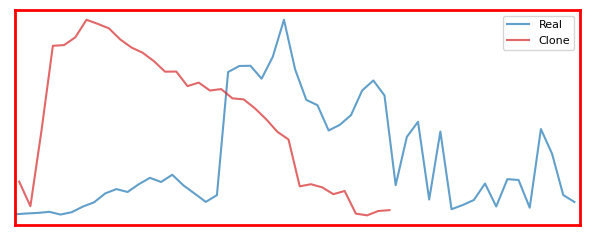}
    \caption{\textcolor{red}{Fraud}}
  \end{subfigure}
  \caption{Temporal segments from the original and synthesized signals. Mispronounced word "fraud" detected.}
  \label{fig:segments}
\end{figure}

The utterance under examination, \emph{“Robbery, Bribery, Fraud,”} is presented in Figure~\ref{fig:phones} with its corresponding phonetic transcription. We extract the relevant temporal segments from both the original recording and the synthesized audio; these segments are depicted in Figure~\ref{fig:segments}. Subsequent alignment is performed via Dynamic Time Warping (DTW), as demonstrated in Figure~\ref{fig:dtw}.


\begin{table}[!t]
  \centering

  \begin{tabular*}{\textwidth}{@{\extracolsep{\fill}} l l *{5}{S} }
    \toprule
    Dataset & Outcome & {Mean} & {Median} & {Std} & {25\%} & {75\%} \\
    \midrule
    \multirow{2}{*}{EBVS} & Correct   & 0.3467 & 0.3343 & 0.0981 & 0.2694 & 0.4114 \\
                          & Incorrect & 0.3785 & 0.3795 & 0.0928 & 0.3130 & 0.4378 \\
    \multirow{2}{*}{ERMS} & Correct   & 0.3649 & 0.3575 & 0.1030 & 0.2903 & 0.4289 \\
                          & Incorrect & 0.3952 & 0.3852 & 0.0971 & 0.3261 & 0.4601 \\
    \multirow{2}{*}{MBMPS}& Correct   & 0.3252 & 0.3152 & 0.0936 & 0.2547 & 0.3809 \\
                          & Incorrect & 0.3379 & 0.3301 & 0.0907 & 0.2751 & 0.3957 \\
    \multirow{2}{*}{NJS}  & Correct   & 0.3693 & 0.3551 & 0.1053 & 0.2936 & 0.4349 \\
                          & Incorrect & 0.3971 & 0.3913 & 0.0960 & 0.3270 & 0.4608 \\
    \bottomrule
  \end{tabular*}
    \caption{Summary of distances across datasets for correct vs.\ incorrect examples.}  \label{tab:summary}
\end{table}


Focusing on the final word \emph{“Fraud,”} the phonetic sequence can be analyzed as follows:
\begin{itemize}
  \item /f/ (voiceless labiodental fricative)
  \item /r/ (voiced postalveolar approximant)
  \item \textbf{Vowel:} original sample (incorrectly realized as /AW1/), whereas the synthesized output correctly produces /AO1/
  \item /d/ (voiced alveolar plosive)
\end{itemize}



\begin{figure}[!t]
  \centering

  \begin{subfigure}[t]{0.4\linewidth}
    \centering
    \includegraphics[width=\linewidth]{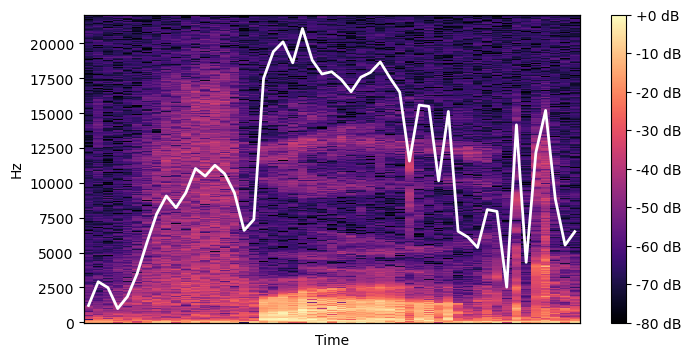}
    \caption{Real-world slice}
    \label{fig:STFT-real}
  \end{subfigure}\;
  \begin{subfigure}[t]{0.4\linewidth}
    \centering
    \includegraphics[width=\linewidth]{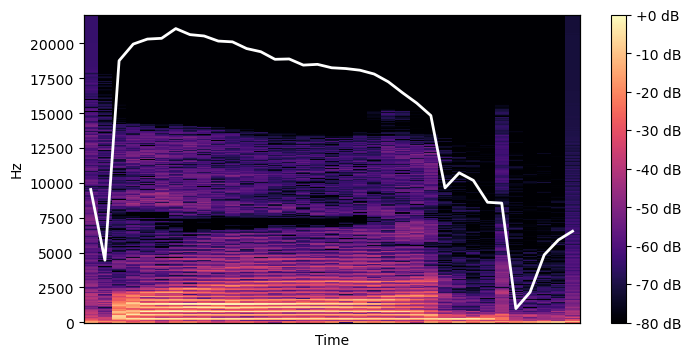}
    \caption{Clone-world slice}
    \label{fig:STFT-clone}
  \end{subfigure}

  \caption{STFT spectrograms of a real vs.\ cloned word slice
           (colour scale: \(-80\,\mathrm{dB}\) to \(0\,\mathrm{dB}\)).
           The white curve shows the frame-wise mean magnitude.}
  \label{fig:STFT}
\end{figure}

Notably, the original speaker’s production employs /AW1/, indicative of the merger, while our model distinguishes the difference with /AO1/. This contrast substantiates the model’s ability to mitigate L1 interference effects by reinstating the phonetic distinction lost in non‐native renditions.


\section{Conclusion and Future Work}
This paper introduces a novel paradigm for pronunciation error detection through voice cloning and differential acoustic analysis, demonstrating robust performance in identifying mispronunciations across diverse linguistic backgrounds. Our experiments validate the core hypothesis that deviations between original and pronunciation-corrected cloned speech correlate with human-perceived errors. While the current framework operates effectively without language-specific rules, future work will integrate linguistic knowledge to enhance error type classification (e.g., distinguishing phonemic substitutions from prosodic errors) and pedagogical feedback. Further developments will focus on real-time implementation for interactive language learning applications and expansion to under-resourced languages, addressing broader cross-lingual pronunciation challenges. By bridging voice cloning technology with language pedagogy, this approach establishes a scalable foundation for adaptive pronunciation training systems that preserve speaker identity while targeting acoustic-phonetic improvements.

\bibliographystyle{IEEEtran}
\bibliography{ref}
\end{document}

%% file: algo_2.tex
\subsection{Algorithm}
\begin{breakablealgorithm}
\caption{Mispronunciation Detection via TTS-Clone Comparison}
\begin{algorithmic}[1]
\Require Real utterances $U=\{u_i\}_{i=1}^M$, TTS model $V$, KDE estimators $K_{\mathrm{COR}}$, $K_{\mathrm{INC}}$, thresholds $\tau_C,\tau_I$
\State \textbf{Training Phase:}
\For{each training utterance $u_i \in U$}
    \State $\hat u_i \leftarrow V(u_i)$ \Comment{Generate TTS clone}
    \State Align words: $\{(w_j, [s_j^{\mathrm{real}},e_j^{\mathrm{real}}], [s_j^{\mathrm{clone}},e_j^{\mathrm{clone}}])\}_j$
    \For{each aligned word $w_j$}
        \State Extract MFCC envelopes $E_j^{\mathrm{real}} \in \mathbb{R}^{13\times T_j^{\mathrm{real}}}$, $E_j^{\mathrm{clone}} \in \mathbb{R}^{13\times T_j^{\mathrm{clone}}}$
        \State Resample to $\hat{T} = \max(T_j^{\mathrm{real}}, T_j^{\mathrm{clone}})$ yielding $\tilde{A}, \tilde{B} \in \mathbb{R}^{\hat{T}\times13}$
        \For{each coefficient $d \in \{1,...,13\}$}
            \State Compute DTW distance $d_d$ between $\tilde{A}_{:,d}$ and $\tilde{B}_{:,d}$
        \EndFor
        \State $\bar{d}_j \leftarrow (13\hat{T})^{-1}\sum_{d=1}^{13} d_d$ \Comment{Normalized distance}
        \State Append $\bar{d}_j$ to $\mathcal{D}_{\mathrm{correct}}$ or $\mathcal{D}_{\mathrm{incorrect}}$ per ground truth
    \EndFor
\EndFor
\State Fit KDE: $K_{\mathrm{COR}} \leftarrow \text{kde}(\mathcal{D}_{\mathrm{correct}})$, $K_{\mathrm{INC}} \leftarrow \text{kde}(\mathcal{D}_{\mathrm{incorrect}})$
\State Compute thresholds: $\tau_C \leftarrow Q_{0.9}(\mathcal{D}_{\mathrm{correct}})$, $\tau_I \leftarrow Q_{0.9}(\mathcal{D}_{\mathrm{incorrect}})$
\State \textbf{Runtime Detection (new utterance $u$):}
\State $\hat u \leftarrow V(u)$
\State Repeat alignment and feature extraction to obtain $\{\bar{d}_j\}$
\For{each word $w_j$ with distance $\bar{d}_j$}
    \State Evaluate KDE: 
    $k_{\mathrm{cor}} \leftarrow K_{\mathrm{COR}}(\bar{d}_j)$, 
    $k_{\mathrm{inc}} \leftarrow K_{\mathrm{INC}}(\bar{d}_j)$
    \State Classify:
    $c(\bar{d}_j) \leftarrow \mathbb{I}(k_{\mathrm{inc}} > k_{\mathrm{cor}})$
    \State Select threshold:
    \[
    \tau^* \leftarrow 
    \begin{cases}
    \tau_C, & \text{if } c(\bar{d}_j) = 0,\\
    \tau_I, & \text{if } c(\bar{d}_j) = 1
    \end{cases}
    \]
    \State Predict:
    \[
    \hat y_j \leftarrow
    \begin{cases}
    \text{CORRECT}, & \text{if } c(\bar{d}_j)=0 \land \bar{d}_j \leq \tau_C,\\
    \text{INCORRECT}, & \text{if } c(\bar{d}_j)=1 \land \bar{d}_j \geq \tau_I,\\
    \text{AMBIGUOUS}, & \text{otherwise}
    \end{cases}
    \]
\EndFor
\end{algorithmic}
\end{breakablealgorithm}